\documentclass[a4paper,fleqn]{cas-sc}

\usepackage[authoryear,longnamesfirst]{natbib}
\usepackage{xcolor} 
\usepackage{subcaption}
\usepackage{bm}
\usepackage{multirow}% http://ctan.org/pkg/multirow
\usepackage{booktabs}% http://ctan.org/pkg/booktabs
\usepackage{float}

\usepackage{lineno}

\begin{document}
\let\WriteBookmarks\relax
\def\floatpagepagefraction{1}
\def\textpagefraction{.001}

\shorttitle{Hilda asteroids' forced plane}    

\shortauthors{Matheson, Malhotra}  

\title [mode = title]{On the forced orbital plane of the Hilda asteroids}  




\author[1]{Ian C. Matheson}[orcid=0000-0003-0940-7176]

\cormark[1]


\ead{ianmatheson@arizona.edu}


\credit{Investigation, Formal analysis, Software, Visualization, Data curation, Writing - original draft, Writing - review \& editing}

\affiliation[1]{organization={Department of Aerospace and Mechanical Engineering, The University of Arizona},%Department and Organization
            addressline={1130 N. Mountain Ave., P.O. Box 210119}, 
            city={Tucson},
            postcode={85721}, 
            state={AZ},
            country={USA}}

\author[2]{Renu Malhotra}
[orcid=0000-0002-1226-3305]
% Footnote of the second author

% Email id of the second author
%\ead{}


\credit{Conceptualization, Funding acquisition, Supervision, Writing - review \& editing}

\affiliation[inst2]{organization={Lunar and Planetary Laboratory, The University of Arizona},%Department and Organization
            addressline={1629 E. University Blvd.}, 
            city={Tucson},
            postcode={85721}, 
            state={AZ},
            country={USA}}


\cortext[cor1]{Corresponding author}


\begin{abstract}
Hilda-group asteroids librate in Jupiter's interior 3:2 mean motion resonance.
We estimate that the Hilda group is observationally complete up to absolute magnitude $H\leq16.3$.
This provides a statistically useful sample of thousands of resonant objects, all within a narrow range of semi-major axes, to compare with theoretical expectations of their orbital distribution from dynamical theory.
We use von Mises-Fisher statistics to calculate the sample mean planes and mean plane uncertainties for the Hilda group and its Hilda, Schubart, and Potomac collisional subfamilies.
Although Laplace-Lagrange linear secular theory is considered inapplicable within mean motion resonances,
we find that the Laplace plane and the orbital plane of Jupiter are both statistically indistinguishable from the sample mean plane of the Hildas.
In future work, we intend to extend this investigation to resonant populations in the Kuiper belt so as to further test the validity of Laplace-Lagrange linear secular theory for the mean planes of resonant populations.
\end{abstract}


\begin{highlights}

\item We find the mean orbit planes of the Hilda group and its collisional subfamilies.

\item They are all statistically indistinguishable from the Jupiter and Laplace planes.

\item We rule out the invariable plane as the mean orbital plane of the Hildas.

\end{highlights}

% Keywords

\maketitle

%\begin{linenumbers}

% Main text
\section{Introduction}
\label{s:introduction}
The mean orbital plane of groups of small bodies in the solar system is understood to be forced by the secular gravitational perturbations of the major planets.
For small bodies in the solar system that are not subject to a planetary mean motion resonance (MMR), the Laplace-Lagrange linear secular theory provides an estimate of the forced orbital plane as a function of semi-major axis, when the small body is treated as a massless test particle \citep{md99}.
For a group of non-resonant small bodies with a small dispersion in semimajor axis, their average forced plane according to linear secular theory provides an excellent theoretical estimate of their mean orbital plane \citep{cc08,vm17,mm23}.

In the present work, we ask: what is the forced orbital plane of dynamical groups of resonant small bodies? 
The Laplace-Lagrange secular theory is ostensibly not applicable in MMR zones, as the resonant terms in the disturbing function are omitted from its derivation, as in \citet[chapters 6-7]{md99}, although this limitation is not well defined.
Typical textbook reviews of resonant small-body behavior, as in \citet[chapter 8]{md99}, focus on the coplanar case and omit out-of-plane motion.
The literature on out-of-plane motion of small bodies near mean motion resonances appears to be silent on the forced plane and focuses on the determination of the ``proper inclination'' as one of the quasi-integrals that are nearly time-invariant in the long term dynamics of small bodies \citep[e.g.][]{Knezevic:2002,Nesvorny:2024}.
Previous studies based on a succession of simplified models for the Hilda asteroids estimated that the coupling between the mean motion resonance and the orbital plane inclination leads to only a small inclination forcing amplitude of not more than a few tenths of a degree \citep{schubart1982numerical,schubart1982three}.
However, those studies did not explicitly probe the forced plane of the Hilda group.

Here we probe the question of the forced plane of a group of small bodies librating in a mean motion resonance.
We use observational data and numerical simulation as the most convenient means of studying this question.
The Hilda asteroids, librating in Jupiter's interior 3:2 MMR, represent a statistically significant sample of nearly 4000 objects which is observationally complete to absolute magnitude $H\leq16.3$ (see Section \ref{s:hildas}), so we can use them as a test case to investigate their mean orbital plane while limiting our uncertainties to those inherent in the sample statistics without concern for observational survey biases.
With this large observational sample, we test the hypothesis that the mean plane of this resonant group of asteroids is given by the forced plane defined by the Laplace-Lagrange secular theory.
We also discuss the invariable plane and Jupiter's orbital plane as possible forced planes.

Previous studies of the Hilda asteroids have measured their reflectance spectra, finding a correlation between magnitude and surface composition (mainly D- and P-type) in the Hilda group \citep{dahlgren1997study} that suggests \citep{wong2017color,wong20170} a possible common origin with the Jupiter trojans, which have a remarkably similar magnitude-color distribution.
\citet{franklin2004hilda} argued that the eccentricity distribution of the observed Hildas is evidence that this primordial population was swept up in Jupiter's interior 3:2 MMR as Jupiter migrated inward.
\citet{brovz2011did} add the requirement for this migration to take place during the Late Heavy Bombardment, in a system with an extra giant planet (later ejected), and with a substantial YORP effect.
Further extending this line of reasoning, \citet{vokrouhlicky2016capture} suggested that the primordial Hilda and Trojan populations were interlopers from the trans-Neptunian region injected inward by the giant planet instability, and that their subsequent depletion has been largely driven by the YORP mechanism.
These holistic studies of long-term Hilda evolution are fairly successful in accounting for the observed composition, size, and eccentricity distributions of the Hilda group asteroids, but these studies have not addressed the relatively simple question of what the forced plane of this resonant population might be.

We present the results of our investigation by organizing the rest of this paper as follows.
\begin{itemize}
\item In section \ref{s:laplace-theory}, we provide a brief description of the Laplace-Lagrange linear secular theory for the forced plane of a minor planet.
This provides theoretical context for the analysis of the observational data.
\item In section \ref{s:mean-plane-statistics}, we briefly describe the von Mises-Fisher (vMF) distribution, which forms the basis of our calculations of the mean plane and its uncertainty, and we justify its application to the mean orbital plane of the Hilda asteroids.
We also describe the statistical method we use to find the mean plane and mean plane uncertainty for the sets of asteroid clones in Section \ref{s:hildas-xcc-integration}.
\item In section \ref{s:hildas}, we present our sample of Hilda-group asteroids and the collisional families within.
\item In section \ref{s:hildas-xcc-today}, we show the present-day mean plane of the Hilda group and its collisional families, and we comment on its relation to the forced plane, the orbital plane of Jupiter, and the invariable plane of the Solar System.
\item In section \ref{s:hildas-xcc-integration}, we comment on the motion of the mean plane over time in relation to the forced plane, for the observed Hilda asteroids and for sets of clones of selected individual asteroids.
\item Section \ref{s:discussion} provides discussion and summary of our work.
\end{itemize}

\section{Laplace-Lagrange Linear Secular Theory}
\label{s:laplace-theory}
In sections \ref{s:hildas-xcc-today}-\ref{s:hildas-xcc-integration}, we will compute the mean plane of the observed Hilda-group asteroids at the present time, then simulate the orbits of the Hildas for several Myr and compute their mean plane over that time.
This section outlines the dynamical theory we will use to interpret the present-day observed mean plane and the computed mean plane in the future.

Laplace-Lagrange linear secular theory is a simplified description of the secular dynamics of massive planets and massless small bodies obtained by expanding the disturbing function to low order and discarding short-period and resonant terms.
This summary of the theory follows \citet[chapter 7]{md99}.

The simplified physical model is that of a single small body, represented as a massless point, and $N$ point-mass planets orbiting a point-mass Sun, ignoring the collective gravitational effects of a massive disk of small particles.
The orbits of the small body and each massive planet are given as heliocentric osculating elements.
The disturbing function for the small body as well as of the planets is expanded to first order in the masses, second order in the eccentricities, and second order in the inclinations.
Short-period and resonant terms are discarded, leaving only the secular terms of the direct part of the disturbing function, as the indirect part has no secular terms.
The absence of resonant terms then implies that the semimajor axes are constant in time.

It is convenient to rewrite the disturbing function 
as a quadratic form in the variables $h=e\sin{\varphi}$, $k=e\cos{\varphi}$, $p=\sin{i}\sin{\Omega}$, and $q=\sin{i}\cos{\Omega}$.
Here $e,\varphi,i,$ and $\Omega$ respectively denote the orbital eccentricity, the longitude of perihelion, the orbital inclination and the longitude of ascending node; the standard reference plane is the J2000 ecliptic. 
In this paper, we refer to $\Omega$ as simply the ``ascending node''.
The lowest-order formulations of Lagrange's equations of motion for $h$, $k$, $q$, and $p$ for each planet and for the small body then become a system of coupled first order linear differential equations with constant coefficient matrices that couple $k$ with $h$ and $q$ with $p$, while leaving $(k,h)$ uncoupled from $(q,p)$.
Because the orbital planes are fully described by $q$ and $p$, we will omit further discussion of $k$ and $h$.

The orientation of the plane of a Keplerian orbit is given by the unit vector $\hat{\mathbf{h}}$ pointing in the direction of the orbital angular momentum vector, given by
\begin{equation}
\label{e:hhat-vec}
\hat{\mathbf{h}}=(h_x,h_y,h_z)=(\sin{i}\sin{\Omega},-\sin{i}\cos{\Omega},\cos{i}),
\end{equation}
which depends only on $i$ and $\Omega$, or in other words on $q$ and $p$.
For convenience, we prefer to express the orbital plane as a unit vector $\hat{\mathbf{q}}$ on the surface of the unit sphere where
\begin{equation}
\label{e:qhat-vec}
\hat{\mathbf{q}}=(-h_y,h_x,h_z)=(q,p,s)=(\sin{i}\cos{\Omega},\sin{i}\sin{\Omega},\cos{i}).
\end{equation}

The coupled first-order linear differential equations in $q$ and $p$ can be solved to find $\mathbf{q}=(q,p)$ as a function of time for the small body and for each of the planets.
For each planet, the solution is a
superposition of $N$ linear modes, one of which has zero frequency and is identified with the invariable plane of the Solar System (the plane normal to the total barycentric orbital angular momentum vector). For the small body, the solution is
given by the sum of a ``free plane'' $\mathbf{q_1}=(q_1,p_1)$ and a ``forced plane'' $\mathbf{q_0}=(q_0,p_0)$.
The free plane $\mathbf{q_1}$ precesses at a fixed rate $B$ around the forced plane $\mathbf{q_0}$, which is itself a weighted superposition of the linear modes of the orbital planes of the planets.

Note that $\mathbf{q_0}$ is defined in Laplace-Lagrange linear secular theory as the forced plane for a small body at a specific semimajor axis, and henceforth we will call it the ``Laplace plane''. 
The total orbital plane $\mathbf{q}=(q,p)$ is given by \citet[eq. 7.146]{md99}:
\begin{align}
q &= q_1(t)+q_0(t)= \sin{i}_\mathrm{free}\cos(Bt+\gamma) + q_0(t), \label{e:q}\\
p &= p_1(t)+p_0(t)= \sin{i}_\mathrm{free}\sin(Bt+\gamma) + p_0(t). \label{e:p}
\end{align}
In this expression, the free inclination $i_\mathrm{free}$ and phase $\gamma$ are constants computed from the initial conditions of the planetary and small-body orbits, and the frequency $B$ is computed from the masses and (constant) semimajor axes of the planets. The forced plane $\mathbf{q_0}=(q_0,p_0)$ is given by \citet[eq. 7.149-7.150]{md99}:
\begin{align}
q_0(t) &= -\sum_{i=1}^N \frac{\mu_i}{B-f_i}\cos\left(f_it+\gamma_i\right), \label{e:q0}\\
p_0(t) &= -\sum_{i=1}^N \frac{\mu_i}{B-f_i}\sin\left(f_it+\gamma_i\right). \label{e:p0}
\end{align}
The frequencies $f_i$ and the amplitudes $\mu_i$ come from the eigenvalues and eigenvectors, respectively, of the coefficient matrices in the coupled differential equations for $q$ and $p$. 
The elements of the coefficient matrices are computed from the masses of the planets and from their osculating semimajor axes, inclinations, and ascending nodes.
These are the same frequencies and amplitudes as are found in the linear modes of the linear secular solutions for the planets.
The phases $\gamma_i$ are found from the initial conditions.
In a more accurate model, such as in a full multi-body numerical simulation, the quantities $B$, $\mu_i$, $f_i$, and $\gamma_i$ all vary slowly, but they are constant in Laplace-Lagrange linear secular theory.
For the purposes of the present investigation, their variation may be neglected over timescales of several Myr.

When speaking of the Laplace plane, we note that the constant parameters in Eqs.~\ref{e:q}--\ref{e:p0} are calculated for some epoch $t=0$.
We therefore distinguish the ``propagated Laplace plane'', calculated at some time $t=\tau$ from the osculating orbits of the planets at the epoch $t=0$ and setting $t=\tau$ in Eq.~\ref{e:q0}--\ref{e:p0}, from the ``instantaneous Laplace plane'', calculated at $t=\tau$ from the heliocentric osculating orbits of the planets at $t=\tau$ in an multi-body numerical simulation and setting $t=0$ in Eq.~\ref{e:q0}--\ref{e:p0}.
In other words, for the propagated Laplace plane, the constants in Eq.~\ref{e:q0}--\ref{e:p0} are calculated from the heliocentric osculating orbital elements at some epoch $t=0$, and are constant only within the approximations of linear secular theory.
When comparing theory with full multi-body numerical simulations, we recognize that the semimajor axes of the planets and the constants in Eq.~\ref{e:q0}--\ref{e:p0} do not in fact remain constant, so we calculate the instantaneous Laplace plane using the instantaneous osculating orbital elements of the massive planets and the instantaneous heliocentric osculating semimajor axis of the minor planet.
Accordingly, the calculations, figures, and tables in this paper exclusively use the \textit{instantaneous} Laplace plane, not the \textit{propagated} Laplace plane.
It is also important to note that we calculate the instantaneous Laplace plane using the instantaneous heliocentric osculating orbital elements of only the outer planets Jupiter, Saturn, Uranus, and Neptune, neglecting perturbations from the inner planets Mercury, Venus, Earth, and Mars.
Instead, we add the inner planet masses to the mass of the Sun and decrease the outer planet mass ratios relative to the Sun proportionally.

A small body's Laplace plane depends only on its semimajor axis, not on the other elements of its orbit.
Therefore, as shown in theory and simulation by \citet{cc08}, linear secular theory makes a simple and testable prediction for the mean orbital plane of a group of massless small bodies under the condition that the object have similar semimajor axes.
Objects with slightly different semimajor axes will precess around nearby, but not identical, Laplace planes at slightly different rates.
Given enough time, their orbital planes form a thin ring in $q$ and $p$.
The forced orbital plane of the objects considered as a group may be identified with the Laplace plane of some appropriate statistical measure of their central semimajor axis, and the mean orbital plane may be identified with some appropriate statistical measure of the center of the orbital planes in $q$ and $p$.
The mean orbital plane should, to within some measure of statistical confidence, coincide with the forced plane of the Laplace-Lagrange theory at the central semimajor axis.

Laplace-Lagrange linear secular theory produces a well-defined forced plane for most semimajor axis locations, but it must be used with caution.
The $f_i$ frequencies are fixed by the planets' parameters, while the $B$ frequency depends both on the planets' parameters as well as on the semimajor axis of the massless small body.
When $B-f_i\rightarrow0$, the coefficients of the $i$th terms of $q_0$ and $p_0$ (Eq.~\ref{e:q0}--\ref{e:p0}) grow without bound, and the forced plane has an unbounded inclination (in the linear approximation).
These singularities are known as ``secular resonances'', and must be analyzed using a higher-order theory to resolve the singularities; 
the first-order results are of limited accuracy near these secular resonances. 
Secular resonances have actually been identified to exist in narrow regions inside Jupiter's 3:2 MMR \citep[e.g.,][]{Morbidelli:1993,Ferraz-Mello:1998}. 
However, the real Hilda asteroids are absent from these regions and are unaffected by these secular resonances \citep{morbidelli1993secular,zain2025collisional,nesvorny1997asteroidal}, so we do not address them further in this paper.

Furthermore, as we noted in section~\ref{s:introduction}, because the first step in the derivation of linear secular theory is the discarding of resonant terms from the disturbing function, the theory is not constructed to be applicable to massless small bodies located in planetary MMRs, and it must certainly be applied with caution nearby.
Nevertheless, in the absence of an alternative, the Laplace-Lagrange theory is useful in our investigation as a comparison for the measured mean plane of the observed Hilda asteroids. 

\section{Statistics of the Mean Plane}
\label{s:mean-plane-statistics}
For an observationally complete sample of a group of $n$ minor planets, their mean plane and its measurement uncertainty can be estimated with a relatively simple statistical approach. 
Estimating the uncertainty of the mean plane requires the greater effort, which we detail in this section. 
Let us assume that the orbital planes of a group of $n$ minor planets, described by their $(q_i,p_i)$ values, form an approximately circular disk in the $(q,p)$ plane, or an annulus of some average radius and narrow width, or an arc of such an annulus.
In the disk and annular cases, we treat the orbital planes as unit vectors and apply von Mises-Fisher statistics on the unit sphere to find the mean plane and its uncertainty. 
We outline these statistics in Section \ref{ss:vmf-distribution}.
For the circular arc case, we use a circle-fitting algorithm to find the center and radius of the best-fit circle.
We outline the circle-fitting algorithm and its uncertainty in Section \ref{ss:circle-fit-mean-plane}.

\subsection{The von Mises-Fisher Distribution}
\label{ss:vmf-distribution}

The von Mises-Fisher distribution is the simplest analogue of the Gaussian normal distribution on the unit sphere.
The probability density function is given by
\begin{equation}
\label{e:vmf-density}
f(\hat{\mathbf{q}};\hat{\mathbf{\mu}},\kappa)=\frac{\kappa}{4\pi\sinh\kappa}\exp\left( \kappa \hat{\mathbf{q}}_0^T \hat{\mathbf{q}} \right),
\end{equation}
where $\hat{\mathbf{q}}$ is the full unit vector for the orbit plane, $\hat{\mathbf{q}}_0$ is the unit vector for the mean plane, and $\kappa>0$ is the concentration parameter.

The vMF distribution is rotationally uniform around $\hat{\mathbf{q}}_0$.
Larger values of $\kappa$ result in distributions that are more highly concentrated around the mean direction.
The mean direction $\bar{\hat{\mathbf{q}}}_0$ of a sample of $n$ orbital planes $\hat{\mathbf{q}}_i$, $i=1,...,n$, is found by adding up the $\hat{\mathbf{q}}_i$ and normalizing to the unit sphere.
The uncertainty in the sample mean direction $\bar{\hat{\mathbf{q}}}_0$ is described by small circles on the unit sphere centered on $\bar{\hat{\mathbf{q}}}_0$.
The angular width $\phi_{95\%}$ that encloses 95\% of the probability mass in the sampling distribution for $\bar{\hat{\mathbf{q}}}_0$ is calculated as described in \citet{mmk23}.

It should be understood that
when the sample mean pole is $\hat{\mathbf{k}}=(0,0,1)$, the projection of the mean plane uncertainty circle on the $(q,p)$ plane is precisely circular, but if the inclination of the sample mean pole is nonzero, this circle will be compressed in the radial direction and form an ellipse.
For low mean inclinations, the eccentricity of this ellipse is small and we neglect it.
In this approximation, the mean plane uncertainties in $q$, $p$, $i$, and $\Omega$ are all equal to $\sigma$.
That is, using vMF statistics and for low inclination mean poles, $\sigma_i\approx\sigma_\Omega\approx\sigma_q\approx\sigma_p\equiv\sigma$.
In the flat plane approximation, $\sigma_q$ and $\sigma_p$ quantify the uncertainty in the location of the mean pole in the horizontal and vertical $q$ and $p$ directions, while $\sigma_i$ quantifies that uncertainty in the radial direction relative to the ecliptic pole and $\sigma_\Omega$ quantifies it in the transverse direction.
Because the vMF uncertainty region is effectively a circle, $\sigma_q$, $\sigma_p$, $\sigma_i$, and $\sigma_\Omega$ all measure its diameter in different directions and are all equal.

\begin{centering}
\begin{figure}
\includegraphics[width=0.7\columnwidth]{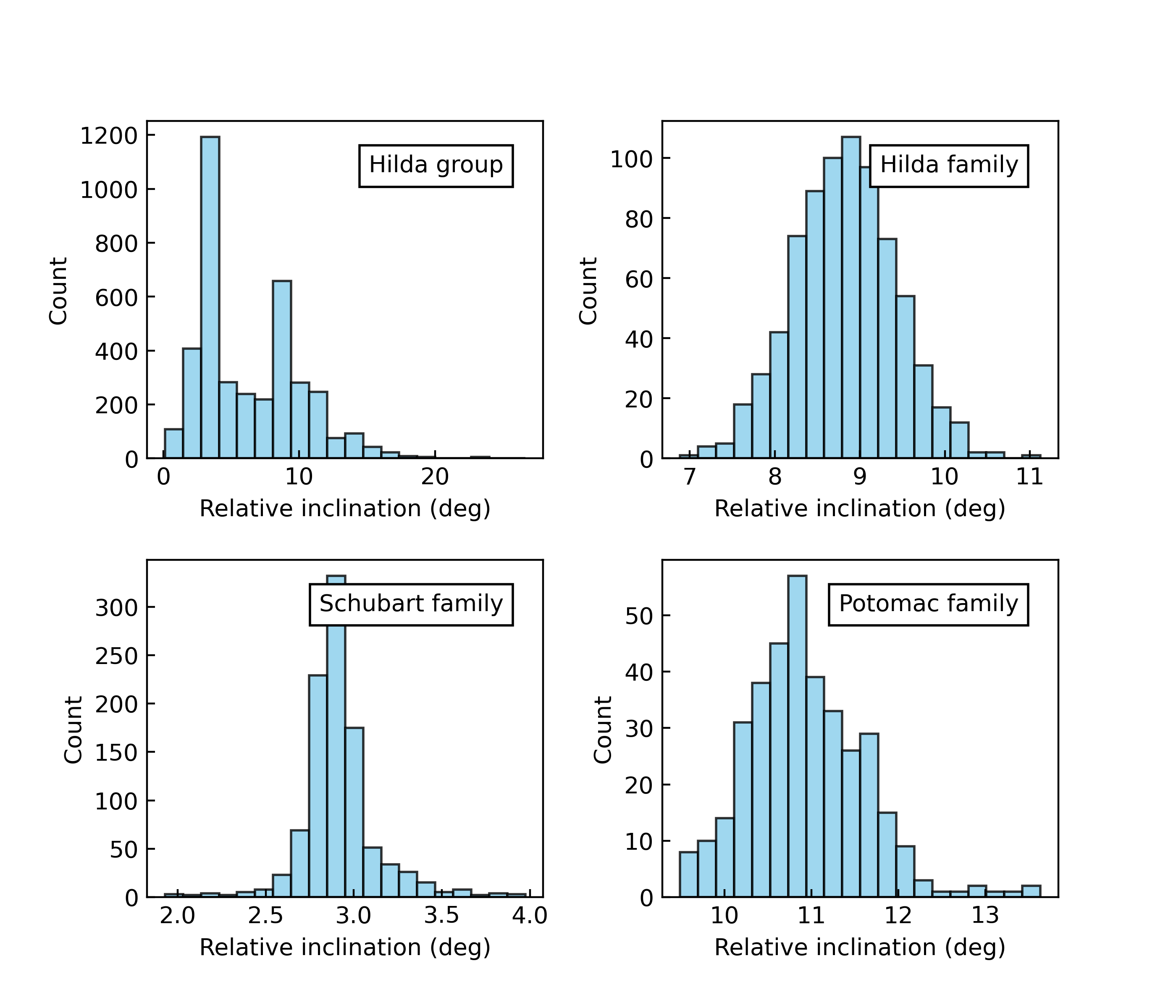}
\caption{
Histograms for the inclinations of the Hilda group, and the Schubart, Hilda, and Potomac collisional families, relative to the vMF mean plane for each population.
}
\label{fig:relative-inclinations}
\end{figure}
\end{centering}

The orbital planes of the Hilda-group asteroids appear rotationally uniform relative to their sample mean plane, but their distribution differs from a vMF distribution (cf. Figure \ref{fig:relative-inclinations}and Figure \ref{fig:plot_qp_zoomout_allgroups_15.69}).
A substantial ``background'' population does resemble a vMF distribution, but the collisional families present as separate concentric thin rings.
Neither the thin rings of the collisional families, nor the Hilda group as a whole, can plausibly be called a vMF distribution.
Nevertheless, we make the ansatz that the sampling distribution of the mean plane does, based on the following reasoning.

First, note that the vMF distribution is obtained by restricting the isotropic uncorrelated trivariate Gaussian distribution to the surface of the unit sphere and recalculating the normalizing coefficient.

Because of this, it is intuitive that the sampling distribution for the mean direction of rotationally uniform distributions on the unit sphere should converge to the vMF distribution, just as the central limit theorem guarantees that the sampling distribution for the mean of general trivariate distributions converges to the Gaussian distribution.
We do not prove this, but we demonstrate it for a few distributions relevant to this work.

As discussed later in section \ref{s:hildas}, the Hilda-group asteroids contain three large collisional families, named after the asteroids Hilda, Schubart, and Potomac.
In this paper, we call the collisional family named for asteroid Hilda the ``Hilda family'', while we call the entire population of 3:2 resonant asteroids the ``Hilda group''.
Those Hilda-group asteroids not associated with any collisional family will be referred to as the ``background Hildas''.
The orbital planes of the Hilda family form a thin ring on the unit sphere with an inclination width of about $9^\circ$.
The orbital planes of the Schubart and Potomac families form thin rings with inclination widths of about $3^\circ$ and $11^\circ$.
The orbital planes of the background Hildas more closely resemble a classic vMF distribution.

To show that the vMF distribution adequately describes the sampling distribution for the mean of each family and of the entire Hilda group, we construct model populations by first computing the vMF mean plane $\bar{\hat{\mathbf{q}}}_0$ of each sample and then computing the inclination for each object in the sample relative to the sample mean.
We draw 100,000 bootstrap samples of relative inclination, pair them with equally sized uniform random samples on the circle for the ascending node, and construct unit vectors.
Next, we compute the vMF sample mean and $\phi_{68\%}$, $\phi_{95\%}$, and $\phi_{99.7\%}$ for each bootstrap sample of unit vectors.
We record how many times the true mean plane $\hat{\mathbf{k}}=(0,0,1)$ is within the 68\%, 95\%, and 99.7\% uncertainty regions.
If the vMF distribution is a suitable description of the sampling distribution of the mean of an arbitrary rotationally uniform distribution on the unit sphere, then the proportion of the time that the true mean plane is within a given uncertainty region should be very close to the percentage prescribed for that uncertainty region.

In Figure \ref{fig:relative-inclinations}, we show histograms of the empirical relative inclination distributions of the Schubart, Hilda, and Potomac collisional families from which our bootstrap samples are drawn, and for the entire Hilda group.
In Table \ref{t:vmf_demo}, we show what proportion of bootstrap sample repetitions yielded sample mean uncertainty regions that enclosed the true mean plane $\hat{\mathbf{k}}$.
These proportions closely match the specified uncertainty levels, showing the vMF distribution to adequately model the distribution of the sample mean for rotationally symmetric samples resembling the orbital planes of the Hilda-group asteroids.
In other words, we have demonstrated that the mean plane obeys vMF statistics even when the sample itself does not, provided the sample is rotationally symmetric. 

\begin{table*}
    \centering
    \begin{tabular}{lrlll}
         Category & $n$ & 68\% & 95\% & 99.7\% \\
         \hline
         All Hildas      & 3893 & 0.68007 & 0.94998 & 0.99700 \\
         
         Schubart family       &  1000 & 0.67912 &  0.94979 & 0.99696 \\
         
         Hilda family    &  757 & 0.67903 &  0.94906 & 0.99688 \\
         
         Potomac family     &  365 & 0.67566 &  0.95000 & 0.99693 \\
    \end{tabular}
    \caption{Columns are category; object count $(H\leq16.3)$; and proportion of bootstrap samples with vMF sample 68\%, 95\%, and 99.7\% uncertainty regions that enclose the true mean plane of the model population, $\hat{\mathbf{k}}$.}
    \label{t:vmf_demo}
\end{table*}

\subsection{The Circle-Fit Mean Plane}
\label{ss:circle-fit-mean-plane}
When the orbital planes of the individual objects form one or more circular arcs in $(q,p)$ without uniformly filling an annulus, it is necessary to use a circle-fitting algorithm to find the center and radius of the best-fit circle.
(We will need this in Section~\ref{s:hildas-xcc-integration} where we employ clones of individual asteroids to investigate the time evolution of their orbital planes.)
There are a variety of circle-fitting algorithms of varying accuracy and stability for short arcs, but for long enough arcs a simple least-squares fit is sufficient.
The Python package \texttt{circle-fit} \citep{circlefit} implements several circle-fitting methods from \citet{al2009error}, but does not compute the covariance matrix for any of them.
We used \texttt{circle-fit} to find the center and radius of each best-fit circle, but we wrote our own Python code to compute its covariance, again following the procedures outlined in \citet{al2009error}, which we reproduce below.
Our code is available on GitHub as described in section \ref{s:data_availability}.

The least-squares circle fit adopts an idealized statistical model in which a model population lies exactly on a circle of center $(\tilde{q}_0,\tilde{p}_0)$ and radius $\tilde{R}$.
The $n$ objects in the model population have model orbital planes $(\tilde{q}_i,\tilde{p}_i)$, related to the model center by
\begin{align}
\tilde{q}_i &= \tilde{q}_0+\tilde{R}\cos{\tilde{\phi}_i}, \\
\tilde{p}_i &= \tilde{p}_0+\tilde{R}\sin{\tilde{\phi}_i}
\end{align}
for some angle $\tilde{\phi}_i$. 
For later convenience, we write 
\begin{align}
\tilde{u}_i=\cos{\tilde{\phi}_i}=\frac{\tilde{q}_i-\tilde{q}_0}{\tilde{R}},\\
\tilde{v}_i=\sin{\tilde{\phi}_i}=\frac{\tilde{p}_i-\tilde{p}_0}{\tilde{R}}.
\end{align}
For the purposes of fitting a circle to the observed orbital planes $(q_i,p_i)$, we assume that the observed planes are the model planes plus random deviations with Gaussian statistics.
That is, the observed orbital planes $(q_i,p_i)$ are related to the model orbital planes $(\tilde{q}_i,\tilde{p}_i)$ as
\begin{equation}
q_i=\tilde{q}_i+\tilde{\delta}_i,\,\,\,\,\,\,\,\,p_i=\tilde{p}_i+\tilde{\varepsilon}_i,
\end{equation}
with $(\tilde{\delta}_i,\tilde{\varepsilon}_i)$ an isotropic bivariate Gaussian random variable with zero mean and a model variance in each direction of $\tilde{\sigma}^2$.

The least-squares circle fit minimizes the sum of squares for the projected distances from the observed points to the model (fitted) circle.
If we write the estimated center and radius as $\hat{\mathbf{\Theta}}=(\hat{q}_0,\hat{p}_0,\hat{R})$ and the model center and radius as $\tilde{\mathbf{\Theta}}=(\tilde{q}_0,\tilde{p}_0,\tilde{R})$, then $\hat{\mathbf{\Theta}}$ is found as
\begin{equation}
\hat{\mathbf{\Theta}}=\mathrm{argmin}\left[\sum_{i=1}^n \left(\sqrt{(q_i-\hat{q}_0)^2+(p_i-\hat{p}_0)^2}-\hat{R}\right)^2\right],
\end{equation}
where the argmin function refers to the values of $(\hat{q}_0,\hat{p}_0,\hat{R})$ that result in the smallest value of $\mathbf{\Theta}$.
The specific numerical method used for the minimization is not important to this paper.
To leading order, the least-squares circle fit has zero bias in $q_0$ and $p_0$.
The small bias in $R$ is not important to this paper, as we are only concerned with the location of the mean plane.
To leading order, the covariance matrix of the fit is 
\begin{equation}
E[(\hat{\mathbf{\Theta}}-\tilde{\mathbf{\Theta}})(\hat{\mathbf{\Theta}}-\tilde{\mathbf{\Theta}})^T]=\tilde{\sigma}^2(\mathbf{\tilde{W}}^T\mathbf{\tilde{W}}^{-1}),
\end{equation}
where $\tilde{\sigma}^2$ is the model variance of the Gaussian noise and 
\begin{equation}
\mathbf{\tilde{W}}=
\begin{bmatrix}
\tilde{u}_1 & \tilde{v}_1 & 1 \\
\vdots & \vdots & \vdots \\
\tilde{u}_n & \tilde{v}_n & 1
\end{bmatrix}.
\end{equation}
Because we do not know the model values of $(\tilde{u}_i,\tilde{v}_i)$ to compute $\mathbf{\tilde{W}}$, we compute an approximate version $\mathbf{\hat{W}}$ using
\begin{align}
\hat{u}_i=\frac{q_i-\hat{q_0}}{\hat{R}},\\
\hat{v}_i=\frac{p_i-\hat{p_0}}{\hat{R}}.
\end{align}
Because we do not know the model value of $\tilde{\sigma}^2$, we compute
\begin{equation}
\hat{\phi}_i=\arctan(p_i-\hat{p}_0,q_i-\hat{q}_0),
\end{equation}
\begin{align}
\hat{q}_i &= \hat{q}_0+\hat{R}\cos{\hat{\phi}_i},\\
\hat{p}_i &= \hat{p}_0+\hat{R}\sin{\hat{\phi}_i},\\
\hat{\delta}_i &= q_i - \hat{q}_i,\\
\hat{\varepsilon}_i &= p_i - \hat{p}_i.
\end{align}
We then calculate an approximate noise variance $\hat{\sigma}^2$ as the variance of the $\hat{\delta}_i$ and the $\hat{\varepsilon}_i$.
This is not rigorously justified, but it aligns with the use of the known sample variance instead of the unknown population variance when calculating confidence intervals for ordinary Gaussian distributions.

Thus our estimated covariance matrix for the circle fit is
\begin{equation}
\hat{E}[(\hat{\mathbf{\Theta}}-\tilde{\mathbf{\Theta}})(\hat{\mathbf{\Theta}}-\tilde{\mathbf{\Theta}})^T]=\hat{\sigma}^2(\mathbf{\hat{W}}^T\mathbf{\hat{W}}^{-1})
\end{equation}
or
\begin{equation}
\hat{E}[\cdot]=
\begin{bmatrix}
\hat{\sigma}_{q_0}^2 & \hat{\rho}_{q_0p_0}\hat{\sigma}_{q_0}\hat{\sigma}_{p_0} &  
\hat{\rho}_{q_0R_0}\hat{\sigma}_{q_0}\hat{\sigma}_{R_0}\\
\hat{\rho}_{q_0p_0}\hat{\sigma}_{q_0}\hat{\sigma}_{p_0} & \hat{\sigma}_{p_0}^2 &  \hat{\rho}_{p_0R_0}\hat{\sigma}_{p_0}\hat{\sigma}_{R_0}\\
\hat{\rho}_{q_0R_0}\hat{\sigma}_{q_0}\hat{\sigma}_{R_0} & \hat{\rho}_{p_0R_0}\hat{\sigma}_{p_0}\hat{\sigma}_{R_0} & \hat{\sigma}_{R_0}^2
\end{bmatrix},
\end{equation}
where $\hat{\sigma}_{q_0}$ is the ordinary Gaussian standard deviation of the estimate for $q_0$, $\hat{\rho}_{q_0p_0}$ is the correlation coefficient between the estimate for $q_0$ and the estimate for $p_0$, etc.

Later, when analyzing the orbit poles of the Hilda, Schubart, Potomac, and Ismene clones, we adopt one-dimensional 95\% confidence intervals for the location of the circle-fit mean plane as $\hat{q}_0\pm1.96\hat{\sigma}_{q_0}$, \,\,$\hat{p}_0\pm1.96\hat{\sigma}_{p_0}$.
This is in accordance with the interval for symmetric 95\% probability mass around the mean of a univariate standard normal distribution.

\section{The Hilda Asteroids}
\label{s:hildas}

The asteroids in Jupiter's interior 3:2 MMR are known as the Hilda group, after asteroid Hilda.
For this study, we used the lists of the 6393 Hilda asteroids and their collisional families provided by \citet{vokrouhlicky2025orbital}.
The asteroids in the lists by \citet{vokrouhlicky2025orbital}
occupy a semimajor axis range of 3.80-4.16 au.
Using code from \citep{hendler2020observational}, we estimate for them an observational completeness limit of absolute magnitude $H\leq16.3$.
That is, to a good approximation, every object with $H\leq16.3$ has been observed.
At the time of this study, the observationally complete sample size is nearly 4000.
This large sample size makes the Hilda group the best candidate for studying the dynamics of the mean plane of a group of resonant objects.
Other potential candidate groups present much smaller observationally complete sample sizes or require a large amount of delicate work to mitigate observational biases to construct a model population.

Within their list of the Hilda group asteroids, \cite{vokrouhlicky2025orbital} reported several collisional families.
They identified these collisional families by using hierarchical clustering analysis of the asteroids' proper elements obtained by means of a fully numerical method; for details, the reader is referred to \citet[][Appendix C]{vokrouhlicky2025orbital}.
%\textcolor{red}
{As alluded to in section \ref{ss:vmf-distribution},
there are three prominent collisional families.}
These are named after the their parent asteroids Hilda, Schubart, and Potomac,
and we call them the ``Hilda family'', ``Schubart family'', and ``Potomac family'' to contrast them with the ``background Hildas'' in no collisional family and the ``Hilda group'' that includes all asteroids in Jupiter's interior 3:2 MMR.
These families present as concentric rings around the forced plane in the space of the osculating elements $(q,p)$.
The Hilda and Schubart families were previously identified by \citet{brovz2011did,schubart1982numerical,schubart1982three}.
There are three other small collisional families  named after their parent asteroids 2008 TG106, Guinevere, and Francette \citep{vokrouhlicky2025orbital}.

Among the 6393 Hilda group asteroids, \citet{vokrouhlicky2025orbital} report 1066 objects in the Hilda family,
1882 in the Schubart family,
506 in the Potomac family,
151 in the Francette family,
54 in the Guinevere family,
and 17 in the 2008 TG106 family.
However, all but one object in the 2008 TG106 family is also in the Schubart family, so we assign the remaining object to the background Hildas and do not consider the 2008 TG106 family any more.

Starting with this dataset, we used Astroquery in Python to download $H$ magnitudes and osculating heliocentric elements in the J2000 ecliptic frame for each object from the JPL Horizons web service \citep{jpl_horizons}.
The orbital elements and $H$ magnitudes used an epoch of JD 2460796.5, or 1 May 2025 00:00:00.
We used the Python code provided by \citet{hendler2020observational} to obtain an observational completeness limit of $H\leq16.3$ for the objects listed by \citet{vokrouhlicky2025orbital}, and then we selected all 3893 objects brighter than that limit.

In our sample of 3893 asteroids with $H\leq16.3$, there are
757 in the Hilda family, 
1000 in the Schubart family, 
365 objects in the Potomac family, 
54 in the Francette and Guinevere families,
and 1717 background Hildas.
Because of the small numbers in the Francette and Guinevere families, we do not discuss these families further in this work.

\begin{figure}
    \centering
    \begin{subfigure}[t]{0.48\textwidth}
        \centering
        \includegraphics[width=\linewidth]{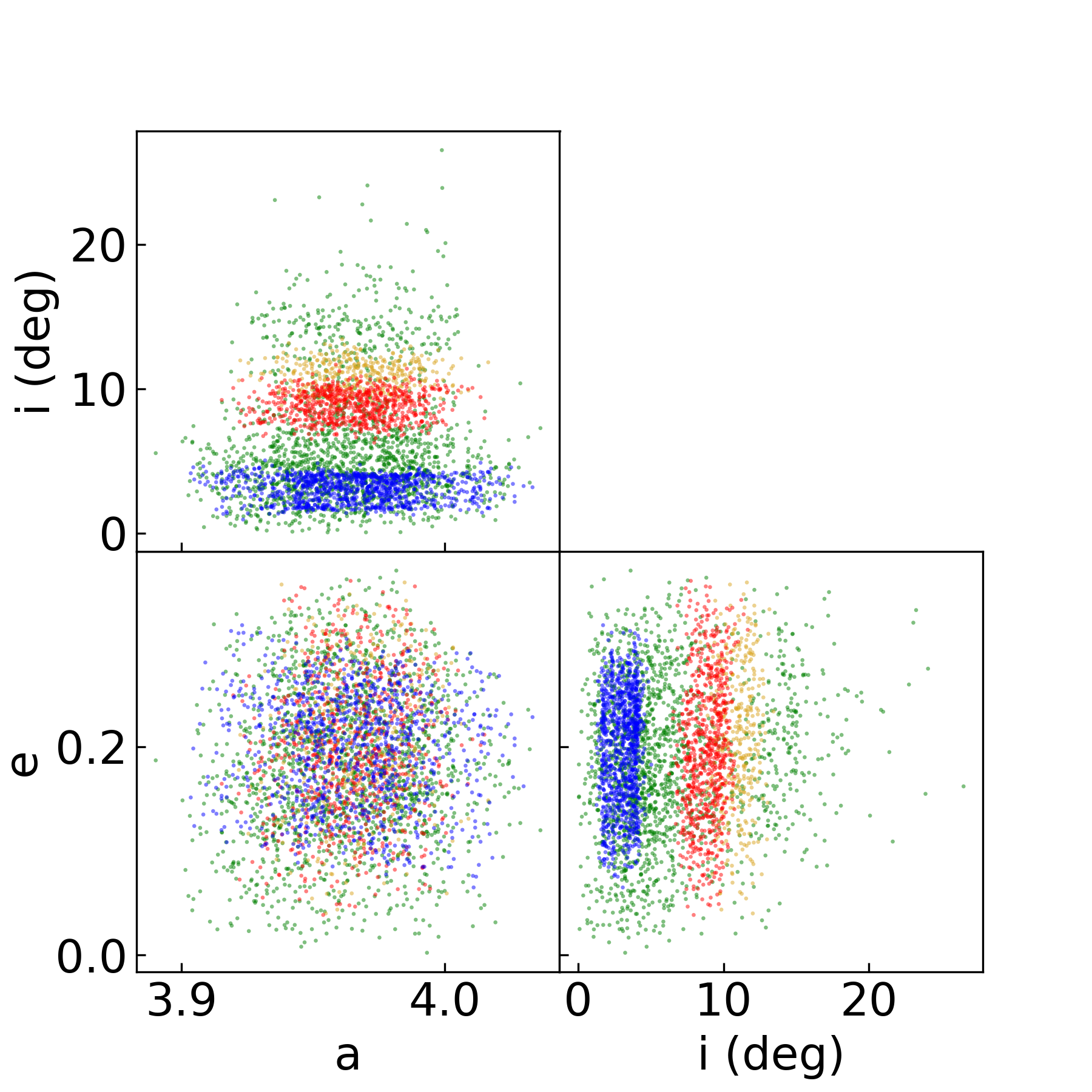} 
        \caption{Semimajor axis, eccentricity, and inclination.
} \label{fig:plot_aei_allgroups_15.69}
    \end{subfigure}
    \hfill
    \begin{subfigure}[t]{0.48\textwidth}
        \centering
        \includegraphics[width=\linewidth]{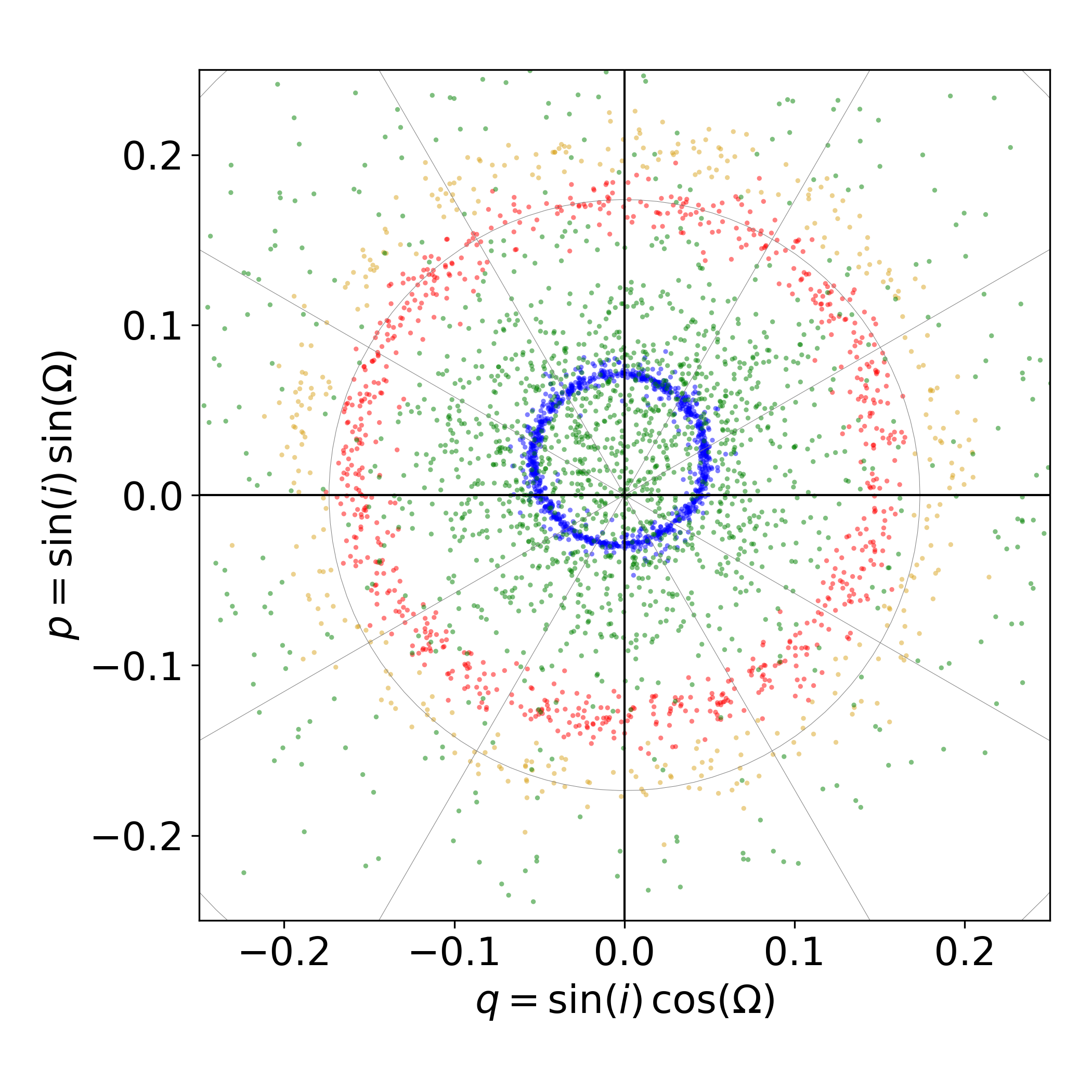} 
        \caption{
        Orbital planes in $q$ and $p$.}

 \label{fig:plot_qp_zoomout_allgroups_15.69}
    \end{subfigure}
\caption{
Heliocentric osculating orbital elements in the J2000 ecliptic frame for the observationally-complete Hilda-group asteroids with absolute magnitude $H\leq16.3$ as of May 1, 2025.
The collisional Hilda, Schubart, and Potomac families are respectively colored red, blue, and goldenrod.
All other objects are colored green.
}
\end{figure}

Figure \ref{fig:plot_aei_allgroups_15.69} shows scatter plots of the osculating heliocentric orbital elements 
as of May 1, 2025 for our observationally complete sample of the Hilda group.
For the group as a whole, we observe that the semimajor axis and eccentricity are uncorrelated.
For each of the collisional families, we observe that the inclination is sharply bounded, and that the semimajor axis and eccentricity are randomly distributed within ranges that are smaller for the higher inclination families. 

The osculating heliocentric orbital planes as of May 1, 2025 are plotted in $q$ and $p$ in Figure \ref{fig:plot_qp_zoomout_allgroups_15.69},
with the Hilda, Schubart, and Potomac families respectively highlighted in red, blue, and goldenrod.
The collisional families appear as three distinct annular rings that are nearly uniformly distributed around their centers.
We will interrogate this uniform appearance later.

It should be noted that collisional families could, in principle, present as an arc if the family were young enough that the orbital planes were not sufficiently differentially precessed to uniformly fill a ring.
This is not the case with the Hilda, Schubart and Potomac families.
According to \citet{milani2017ages}, the Schubart family is roughly 1.6 Gyr old, while the Hilda family is about 5.0 Gyr old.
The timescale to turn a short arc into a nearly uniform ring depends inversely on the initial semimajor axis dispersion.
For reference, we find from numerical integrations that the timescale for a young collisional family to differentially precess and fill out a ring in $q$ and $p$ is a few hundred kyr, if each object in the family begins with the same orbital plane and an osculating semimajor axis dispersion of $\pm$0.01 au from the parent body.
A wider semimajor axis dispersion of $\pm$0.1 au, similar to that observed for the collisional families, can be expected to fill out a ring in $q$ and $p$ in a shorter time, on the order of tens of kyr. 

\begin{centering}
\begin{figure}
\includegraphics[width=0.8\columnwidth]{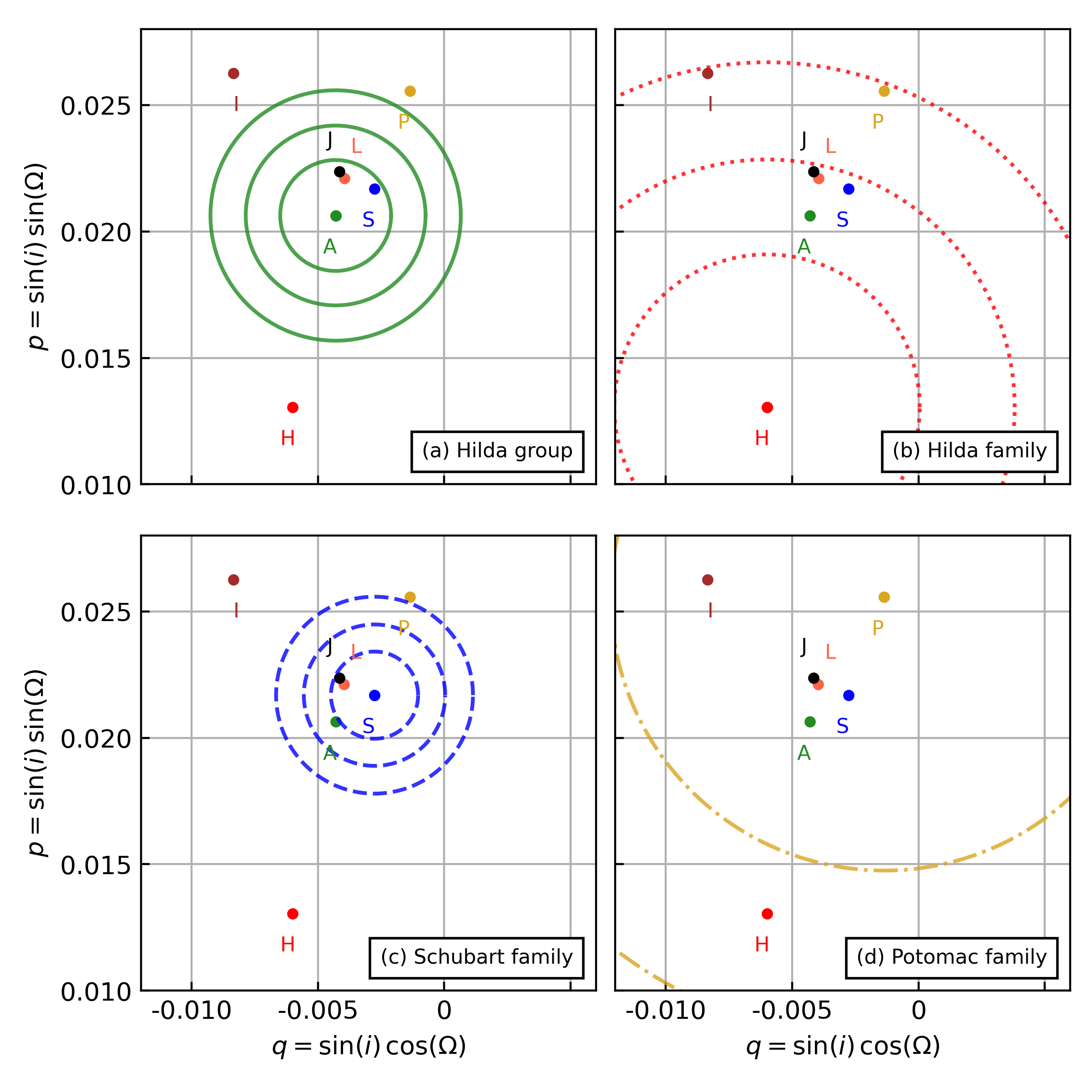}
\caption{
The present-day sample mean planes and their uncertainty regions.
Each subplot contains the invariable plane (purple, I), the Jupiter plane (black, J), the Laplace plane (orange, L), the Hilda group mean plane (green, A), the Hilda family mean plane (red, H), the Schubart family mean plane (blue, S), and the Potomac family mean plane (goldenrod, P), as well as the 68\%, 95\%, and 99.7\% uncertainty regions for the sample mean of the population labeled in the bottom-right corner.
Note that this figure is contained well inside the ring of blue dots in Figure \ref{fig:plot_qp_zoomout_allgroups_15.69}, and this figure does not show the orbital planes of individual asteroids.
In subplot (d), only the 68\% and 95\% uncertainty regions are shown.
}
\label{fig:mean_pole_confidence_overlaps_15.69}
\end{figure}
\end{centering}

\section{The Mean Plane of the Hildas Today}
\label{s:hildas-xcc-today}

Following the methods described in Section \ref{ss:vmf-distribution}, we computed the sample mean plane $\bar{\hat{\mathbf{q}}}_0$ and its uncertainty angles $\phi_{68\%}$, $\phi_{95\%}$, and $\phi_{99.7\%}$ for 
the Hilda group as a whole, and for the collisional Hilda, Schubart, and Potomac families taken separately.
Figure \ref{fig:mean_pole_confidence_overlaps_15.69} displays the sample mean planes and their 68\%, 95\%, and 99.7\% confidence regions for each of the three groups on May 1, 2025.

In these plots, we label the sample mean planes $\bar{\hat{\mathbf{q}}}_0$ of the Hilda, Schubart, and Potomac families as ``H'', ``S'', and ``P'', and we label the sample mean plane of the entire population with $H\leq16.3$ as ``A'' for ``all''.
We label the Laplace plane $\hat{\mathbf{q}}_0$ computed at the mean semimajor axis of the sample as ``L'' for ``Laplace''.
We label the orbital plane of Jupiter as ``J'', and we label the invariable plane of the Solar System as ``I''.
We calculate this by normalizing the total barycentric orbital angular momentum vector of the Sun and planets in our simulation. Here its location in the heliocentric J2000 reference frame is $q=-0.00834$, $p=0.0262$, which is $i=1.58^\circ$, $\Omega=107.6^\circ$.

\begin{itemize}
\item 
Subplot (a) shows the mean plane uncertainty for the entire Hilda group.
The Jupiter plane, forced plane, and Schubart plane are within the 68\% region, while the Hilda family plane, Potomac plane, and invariable plane are outside the 99.7\% region.

\item 
Subplot (b) shows the mean plane uncertainty for the Hilda family.
The Jupiter plane, forced plane, Hilda group plane, and Schubart plane are within the 95\% region, and the Potomac plane and invariable plane are within the 99.7\% region.

\item 
Subplot (c) shows the mean plane uncertainty for the Schubart family.
The Jupiter plane, forced plane, and Hilda group plane are within the 95\% region, and the Hilda family plane, Potomac plane, and invariable plane are outside the 99.7\% region.

\item 
Subplot (d) shows the mean plane uncertainty for the Potomac family.
All other planes are within the 95\% region.

\end{itemize}

We call two planes statistically indistinguishable if either plane is within the 95\% uncertainty region of the other plane.
Figure \ref{fig:mean_pole_confidence_overlaps_15.69} shows that it is statistically plausible that each collisional family has the same mean plane as every other collisional family and the entire Hilda group, and we have no dynamical reason to suspect otherwise.
In each case, the mean plane is statistically indistinguishable from the Laplace plane or from the orbital plane of Jupiter, which is nearly identical to the Laplace plane.
However, the Hilda-group mean plane and the Schubart-family mean plane are statistically distinct from the invariable plane of the Solar System.
Dynamically speaking, if we do not identify the Hilda-group or Schubart-family mean planes with the invariable plane, we have no reason to identify the Hilda-family or Potomac-family mean planes with the invariable plane either.
The larger uncertainty regions for the Hilda-family and Potomac-family mean planes are due to the higher average relative inclinations of the Hilda and Potomac families, as shown in Figure \ref{fig:relative-inclinations}.

The numerical values of the current mean planes of each population are given in Table \ref{t:meanpoles_circlefit}.
The mean inclination for the Hilda family of $i_0=0.82^\circ\pm0.56^\circ$ compares favorably with the \citet{vinogradova2015} result of $i_0=1.20^\circ\pm0.05^\circ$, but the mean longitude of ascending node $\Omega_0=114.66^\circ\pm0.56^\circ$ is far from the \citet{vinogradova2015} result of $\Omega_0=99^\circ\pm1^\circ$. 
\begin{table*}
    \centering
        \begin{tabular}{lrrrr}
         Category & $n$ & $i_0$ & $\Omega_0$ & $\phi_{95\%}$ \\
         \hline
         Hilda group & 3893 & $1.21^\circ$ & 101.$76^\circ$ & $0.20^\circ$ \\
         Hilda family & 757 & $0.82^\circ$ & 114.$66^\circ$ & 0.5$6^\circ$ \\
      Schubart family & 1000 & 1.2$5^\circ$ & 97.2$5^\circ$ & $0.16^\circ$ \\
      Potomac family  & 365 & 1.4$7^\circ$ & $93.04^\circ$ & 1.0$0^\circ$
    \end{tabular}
    \caption{Mean plane locations of the Hilda group, Hilda family, Schubart family, and Potomac family, in degrees, relative to the ecliptic plane.
    Columns are category; object count $(H\leq16.3)$; inclination and ascending node of the mean plane; and the half-angle $\phi_{95\%}$ for the 95\% uncertainty region.}  
    \label{t:meanpoles_circlefit}
\end{table*}

\section{The Mean Plane of the Hildas Over Time}
\label{s:hildas-xcc-integration}

We have shown that, at the present time, the mean planes of the Hilda group and its three collisional families are statistically indistinguishable from each other, and their mean plane is statistically indistinguishable from the Laplace plane or the orbital plane of Jupiter.
We have also shown that the invariable plane is statistically distinct from the mean plane of the Hilda group or the Schubart family.
However, it is conceivable that this observation at the present time is a statistical fluctuation rather than the true state of things.
To show that this observation is dynamically meaningful, we must see whether it persists over secular timescales, and we must check that it applies to individual objects as well as to observed populations.

We used the \textsc{Whfast} integrator in \textsc{Rebound} \citep{rebound} to integrate the giant planets and the Hilda-group asteroids for 2 Myr, with a step size of 0.2 yr and an output cadence of 200 yr.
The solar mass in these integrations was augmented by the total mass of the terrestrial planets, and the mass ratios of the giant planets relative to the Sun were reduced proportionately.
We also chose one member of each population to examine individually, generated 500 clones based on each of the chosen objects, and integrated those clones for 2 Myr with the same step size and output cadence.
We chose asteroid Hilda to represent the Hilda family, asteroid Schubart to represent the Schubart family, asteroid Potomac to represent the Potomac family, and asteroid Ismene to represent Hilda-group asteroids in no collisional family.
Each clone used the orbital elements of the observed asteroid, plus a semimajor axis displacement drawn from the uniform random distribution on $[-0.01,0.01]$ au.
Our use of a uniform initial semimajor axis distribution for the clones, with constant initial inclination and eccentricity, is justified by the uniform semimajor axis distributions of the collisional families seen in Figure \ref{fig:plot_aei_allgroups_15.69}, together with the lack of correlation between semimajor axis and eccentricity.
The width of the semimajor axis dispersion was chosen to keep all clones well within the semimajor axis range of Jupiter's interior 3:2 MMR.
We found by numerical experimentation that increasing the semimajor axis dispersion of the clones by an order of magnitude could result in some clones exhibiting chaotic evolution, which is unsuitable for our purpose.

\begin{centering}
    \begin{figure*}
        \includegraphics[width=\textwidth]{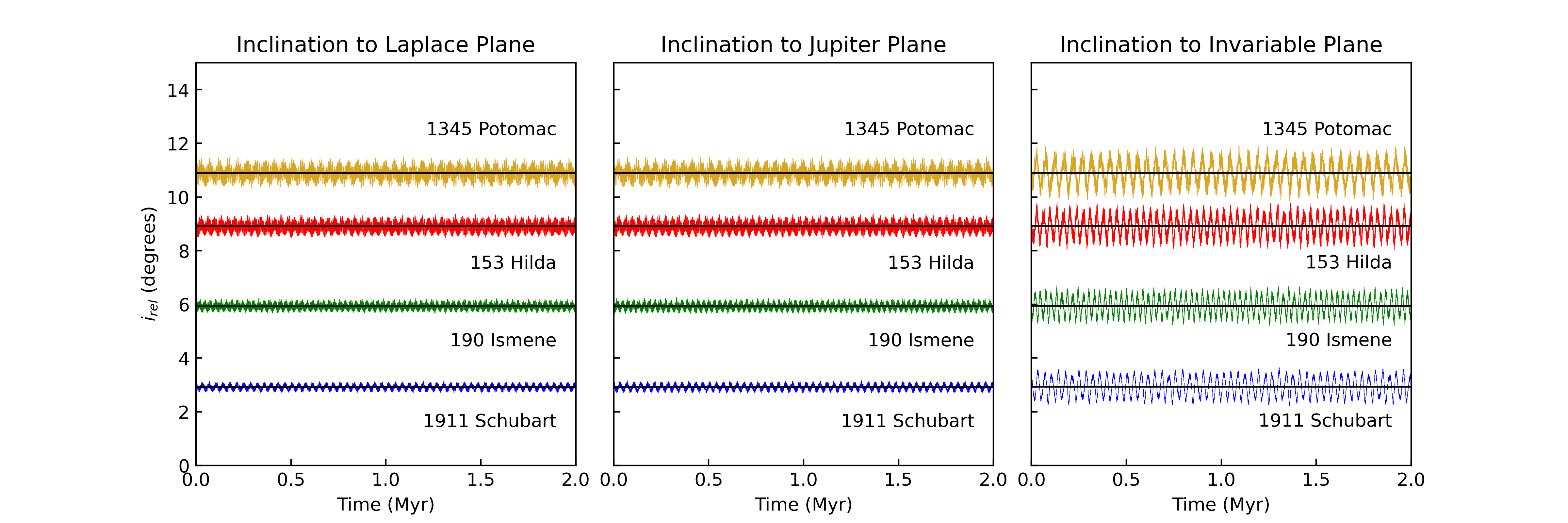}
        \caption{The time-varying angular separation, $i_\mathrm{rel}$, between the osculating orbit plane of the individual asteroids 1911 Schubart, 190 Ismene, 153 Hilda and 1345 Potomac, and the Laplace plane, Jupiter's orbital plane, and the invariable plane. The straight horizontal lines (in black) indicate the time-averaged values for each object.}
\label{fig:fig_irel_plots}
    \end{figure*}
\end{centering}

\begin{centering}
\begin{figure*}
\includegraphics[height=8in]{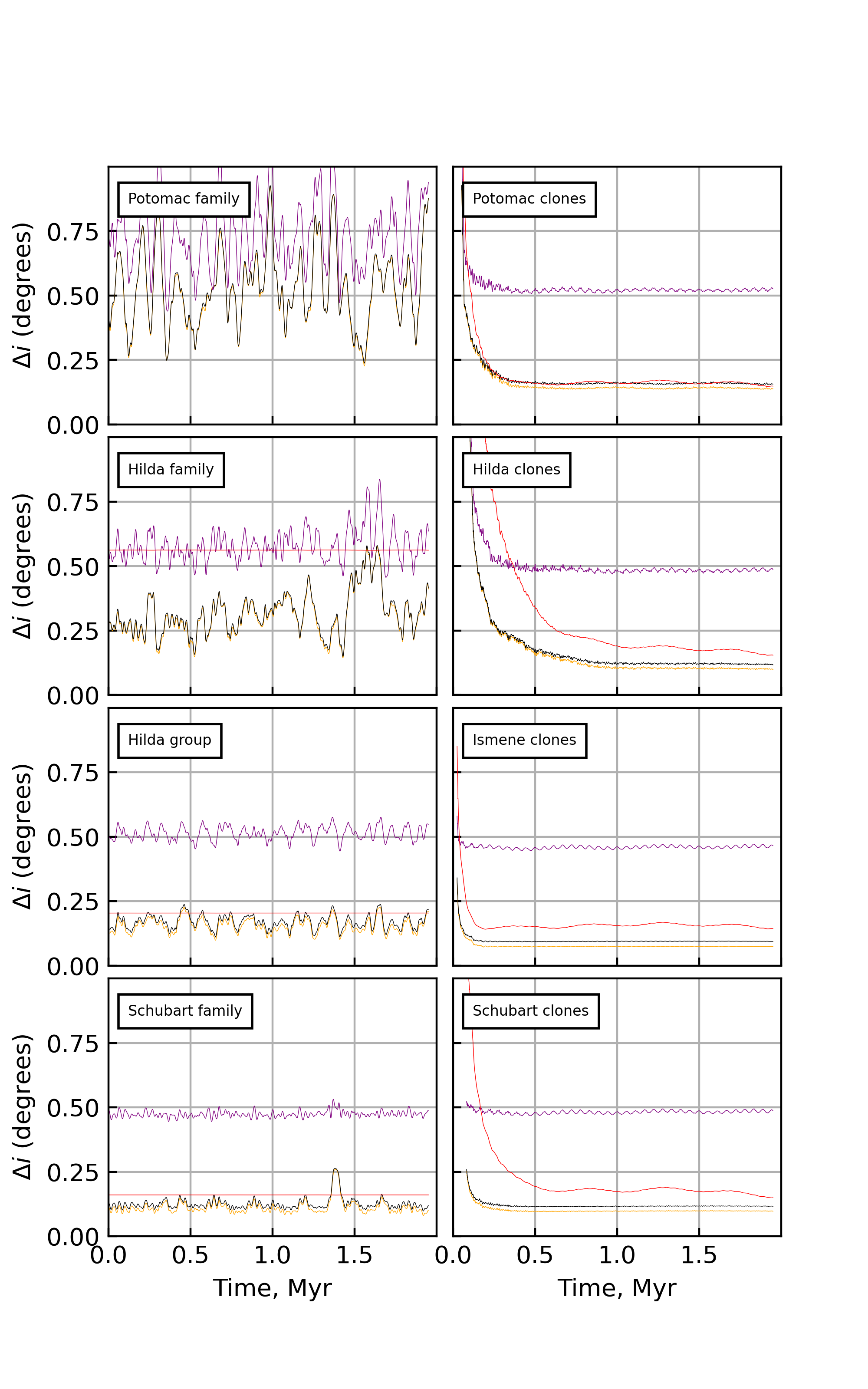}
\caption{
The time-varying angular separation, in degrees, between the mean planes of the indicated groups in each subplot and the invariable plane (purple), Jupiter's (black), and the instantaneous Laplace plane (orange). 
The red trace is the inclination width of the 95\% confidence interval for the mean plane.
To make the differences between the traces easier to see, each trace has been smoothed over a 50-kyr moving average.
}
\label{fig:fig_prm_plots}
\end{figure*}
\end{centering}

The results of these integrations are tabulated in Table \ref{t:delta-incs} and illustrated in  \ref{fig:fig_irel_plots} and Figure \ref{fig:fig_prm_plots}.

In Figure 4, we plot the time-varying angular separation, $i_\mathrm{rel}$, between the osculating orbital plane and the Laplace plane, the orbital plane of Jupiter, and the invariable plane for the four representative individual asteroids, 1911 Schubart, 190 Ismene, 153 Hilda and 1345 Potomac, ordered according to the magnitude of their time-averaged values. 
We find that $i_\mathrm{rel}$, computed relative to the Laplace plane, is conserved to better than about $\pm0.5$ degrees over the 2 Myr integrations for 1911 Schubart, 190 Ismene, and 153 Hilda, and to better than about $\pm0.65$ degrees for 1345 Potomac.
We can see in Table \ref{t:delta-incs} that there are similar values of the time-averaged $i_\mathrm{rel}$ with respect to Jupiter's orbital plane, but there are noticeably larger deviations from the mean with respect to the invariable plane. 
These results indicate that, over time, the individual objects also follow the Laplace plane (and Jupiter's orbital plane) better than they do the invariable plane.

In Table~\ref{t:delta-incs} and Figure~\ref{fig:fig_irel_plots}  we also observe that the magnitude of the deviations from the mean is larger for the higher inclination objects, and that these deviations appear to be dominated by two periods, a shorter period of about 1.5 kyr and a longer period of about 50 kyr, of similar amplitude.
Identifying the source of these small quasi-periodic variations is beyond the scope of the present work, but will be investigated in future work.

\begin{table*}[h!]
    \centering
    \begin{tabular}{l@{\qquad}|rr@{\qquad}|rr@{\qquad}|rr|}
    \multirow{2}{*}{\raisebox{-\heavyrulewidth}{Object}} & \multicolumn{2}{r}{wrt Laplace plane}| & \multicolumn{2}{r}{wrt Jupiter plane}| & \multicolumn{2}{r}{wrt invariable plane}| \\
    \cmidrule{2-7}
    & $\langle i_\mathrm{rel} \rangle$ & $\Delta i$ & $\langle i_\mathrm{rel} \rangle$ & $\Delta i$ & $\langle i_\mathrm{rel} \rangle$ & $\Delta i$ \\
    \midrule
    1911 Schubart   &  2.917 & 0.226 &  2.917 & 0.246 &  2.937 & 0.710 \\
    190 Ismene      &  5.934 & 0.301 &  5.934 & 0.321 &  5.942 & 0.758 \\
    153 Hilda       &  8.916 & 0.456 &  8.916 & 0.475 &  8.922 & 0.895 \\
    1345 Potomac    & 10.891 & 0.610 & 10.891 & 0.628 & 10.897 & 1.031 \\
    \end{tabular}
    \caption{Time-averaged inclinations $\langle i_\mathrm{rel} \rangle$ and maximum inclination variation $\Delta i=\mathrm{max}(\mathrm{abs}(i-\langle i_\mathrm{rel} \rangle))$, in degrees,} of four representative asteroids with respect to the Laplace plane, the Jupiter plane, and the invariable plane. 
    \label{t:delta-incs}
\end{table*}

Figure 5 shows the time-varying angular distance, in degrees, between the sample mean planes of the various populations and the Laplace plane (orange), the orbital plane of Jupiter (black), and the invariable plane (purple).
The red trace is the inclination width of the 95\% confidence interval for the sample mean plane.
The left column (top to bottom) shows this distance for the Potomac collisional family, the Hilda collisional family, the entire Hilda group and the Schubart collisional family; and the right column shows it for the clones of the individual asteroids, Potomac, Hilda, Ismene and Schubart.
In both columns, the data has been smoothed over a moving average of 50 kyr to make the differences between the traces easier to see.
In each case, the invariable plane is substantially farther from the sample mean plane than the Jupiter plane is, and the Jupiter plane is slightly farther from the sample mean than the Laplace plane is.

\section{Summary and Discussion}
\label{s:discussion}
The present investigation was motivated by curiosity to know the forced orbital plane of resonant groups of small bodies in the solar system, as this is not addressed in the literature.
We chose the observationally complete $(H\leq16.3)$ set of Hilda asteroids for this investigation.
This dynamical group librates in Jupiter's interior 3:2 mean motion resonance, and its sample size is nearly 4000.

From the observational data, we measured the mean planes of the Hilda group and the collisional families therein, as well as the uncertainties of these measurements. 
Table \ref{t:meanpoles_circlefit} shows that the orientation of the present-day mean plane of the Hilda group can be measured to an accuracy of $0.2^\circ$, and the mean planes of the Hilda family, Schubart family, and Potomac family can respectively be measured to accuracies of $0.56^\circ$, $0.16^\circ$, and $1.00^\circ$.

We found that the 95\% uncertainty region of the sample mean plane of the Hilda group encompasses both Jupiter's current orbital plane as well as the instantaneous forced plane predicted by the Laplace-Lagrange linear secular theory for the average semimajor axis of the Hilda group,
but it does not enclose the invariable plane.

Furthermore, with $N$-body numerical simulation for 2 Myr, we found that the motion of the sample mean plane of the Hilda group
continues to track the Laplace plane, but not the invariable plane.
Because the Laplace plane in the Hilda region is nearly identical to the Jupiter plane, we cannot distinguish between the Laplace plane and the orbital plane of Jupiter as hypotheses for the true forced plane of the Hilda asteroids.

We tentatively qualitatively explain this result as follows.
Because the Hilda group is in closer proximity to Jupiter and the other giant planets are much farther away and of lower mass, the dynamical environment can be closely approximated by the restricted three-body problem with the Sun, Jupiter, and a massless small body.
In the restricted three-body problem, the instantaneous forced plane is expected to be identical to the Jupiter plane; there is no other preferred plane around which a massless small body's orbital plane can precess.
In the less restricted problem with additional planets on non-coplanar orbits, there are four possible hypotheses for the forced plane: the invariable plane, Jupiter's orbital plane (if Jupiter dominantly controls the dynamics of the resonant population), the Laplace plane, or one that is distinct from all of these if the mean motion resonance strongly affects the forced plane dynamics of resonant populations.
Our results rule out the invariable plane as the forced plane of the Hilda group at high statistical confidence.
However, the present-day observationally complete sample size cannot distinguish between the second and third of these hypotheses for three reasons: the size of the sample, its significant inclination dispersion, and the nearness of the Laplace plane to the Jupiter plane.

We stress that our results for the mean plane of the Hildas should not be extrapolated to other resonant populations of small bodies without further investigation of the specific circumstances. The proximity of the Hildas to Jupiter may be the most significant factor in determining their mean plane, whereas other resonant populations, such as those in Neptune’s exterior mean motion resonances, may have their own peculiar dynamics. We hope to investigate those cases in a forthcoming study.

Our study's main weakness is
the lack of a rigorous procedure to quantify the statistical plausibility of a reference plane as the true forced plane over time, as opposed to its statistical plausibility as the true mean plane at any instant of time.
Because the separation between the instantaneous Laplace plane and  Jupiter's orbital plane is so small, it will be necessary to dramatically increase the sample size to clearly distinguish between the two of them as potential true forced planes for the Hilda asteroids.
In the last five years the observational completeness limit for the Hilda group has been pushed from $H\leq15.7$ \citep{hendler2020observational} to $H\leq16.3$, increasing the observationally complete sample size from 2111 to 3893.
In the future the Vera C. Rubin Observatory and other powerful telescopes may push the observational completeness limit to fainter magnitudes, increasing the sample size enough to permit a more precise measurement of the mean plane of the Hilda asteroids and a more sensitive investigation of its dynamics.
Future studies might also investigate whether unseen distant perturbers can warp the linear secular forced plane enough to account for the time-varying differences between it and the mean plane of the Hilda group, but we have little \textit{a priori} reason to expect that a large, distant unseen planet would have a dramatic effect on the orbital planes of small bodies so close to Jupiter.

\section{Data Availability}
\label{s:data_availability}
The initial orbital elements for the planets and Hilda asteroids, as well as our Python code, are available for download from \url{https://github.com/iwygh/mm26_hildas}.

\section{Acknowledgments}
Funding for IM was provided by NASA FINESST grant 80NSSC23K1362.
We gratefully acknowledge the thoughtful reviews from Matt Clement and the anonymous reviewer.
\printcredits

\bibliographystyle{cas-model2-names}

\bibliography{refs_for_icarus}

%\newpage

%\appendix
%\section{Appendix}\label{sec:sample:appendix} 

%\end{linenumbers}
\end{document}